\documentclass[pdflatex,sn-mathphys-num]{sn-jnl}

\usepackage{graphicx}%
\usepackage{multirow}%
\usepackage{amsmath,amssymb,amsfonts}%
\usepackage{amsthm}%
\usepackage{mathrsfs}%
\usepackage[title]{appendix}%
\usepackage{xcolor}%
\usepackage{textcomp}%
\usepackage{manyfoot}%
\usepackage{booktabs}%
\usepackage{algorithm}%
\usepackage{algorithmicx}%
\usepackage{algpseudocode}%
\usepackage{listings}%

\theoremstyle{thmstyleone}%
\theoremstyle{thmstyletwo}%
\theoremstyle{thmstylethree}%
\usepackage{braket}%
\usepackage{subcaption}%
\usepackage{float}%

\begin{document}

\title[Article Title]{A Unified Quantum Interferometric Framework for Interaction-Free Measurement and Delayed-Choice Experiments}

\author[1]{\fnm{Hitesh} \sur{Awasthi}}\email{hitesh.awasthi.in@gmail.com}
\author[2]{\fnm{Sandeep} \sur{Mishra}}\email{sandeep.mtec@gmail.com}
\author*[3]{\fnm{Anirban} \sur{Pathak}}\email{anirban.pathak@gmail.com}

\affil[1,2,3]{\orgdiv{Department of Physics and Materials Science \& Engineering}, \orgname{Jaypee
Institute of Information Technology}, \orgaddress{\street{A--10, Sector--62}, 
\city{Noida}, \postcode{201309}, \state{Uttar Pradesh}, \country{India}}}

\abstract{A theoretical unified interferometric framework is developed for interaction-free measurement (IFM) and delayed-choice (DC) experiments. The proposed model leads to a quantum circuit based architecture in which an ancillary qubit coherently controls the interaction between a photon and a bomb, allowing the system to evolve in a superposition of interaction and non-interaction regimes. The ancillary control qubit serves as a quantum switch that continuously interpolates between IFM and DC behavior, revealing a common operational origin for both phenomena. We demonstrate that both the phenomena can be realized within this framework through controlled quantum-gate operations. The analysis shows that the framework enables a continuous transition between the particle-like and wave-like behavior in DC setup. The framework is shown to be consistent with standard quantum mechanics, preserving unitary evolution, quantum superposition and the measurement postulate.}


\maketitle

\section{Introduction}\label{sec1}

One of the central features of quantum mechanics is wave-particle duality \cite{debroglie1925}, which challenges classical intuition. This concept is formalized in Bohr's complementarity principle \cite{Bohr1928, Heisenberg1927}, which states that wave and particle aspects of a quantum system are mutually exclusive within a single experimental measurement configuration, thus cannot be observed simultaneously. Quantitatively, however, partial wave and particle characteristics can coexist, as captured by the Englert–Greenberger–Yasin relation \cite{englert1996fringe}. Further,  Mach-Zehnder Interferometer (MZI) has been widely employed to realize and analyze several wave-particle duality experiments, including the interaction-free measurement (IFM) \cite{elitzur1993quantum,kwiat1995interaction}, counterfactual quantum cryptography \cite{GUOSHi, NOH,rao2024protocols}, delayed-choice (DC) experiments \cite{WHEELER19789,ma2016delayed,qureshi2021delayed}, and quantum eraser experiments \cite{Scully,ma2013quantum,qureshi2020demystifying}. The concept of IFM, also known as a counterfactual measurement, was first introduced by Elitzur and Vaidman  through a well-known ``bomb tester" thought experiment \cite{elitzur1993quantum}. Building on this foundational concept Kwiat et al. \cite{Kwiat} enhanced the efficiency of IFM through the quantum Zeno effect \cite{zeno}, thereby paving the way for more practical realization of counterfactual phenomena and extending their application in quantum information theory and communication. In parallel, quantum key distribution (QKD) developed independently with introduction of some famous protocols such as BB84 protocol \cite{bennett1984}, B92 protocol \cite{bennett1992}, E91 protocol \cite{E91} and Goldenberg-Vaidman (GV) protocol \cite{GV} and many more (see \cite{shenoy2017quantum,pirandola2020advances} and references therein).
The advent of IFM introduced the possibility of sharing secret information between two parties without direct particle transmission through the channel  \cite{GUOSHi}. A major advance in counterfactual QKD occurred with the N09 protocol \cite{NOH}. This protocol provides a method in which the shared secret key bits are obtained from detection events without the information carrying particle traversing the quantum channel. Building on this, several works have been done to incorporate the application of IFM in quantum communication \cite{ren2011experimental,Brida_2012,Salih,rao2024protocols,vaidman2019analysis}.

Further, if we look into the roots of DC experiments, then it can be traced back to early discussions on the nature of wave particle duality \cite{debroglie1925}. Within the Copenhagen interpretation, Bohr's principle of complementarity \cite{Bohr1928, Heisenberg1927} states that quantum system exhibit both wave-like and particle-like behavior. To investigate these issues, Wheeler proposed the DC thought experiment \cite{WHEELER19789} designed to test whether a photon decides beforehand to behave wave-like or particle-like or whether the observed behavior is determined only by the final measurement arrangement. The experiment is realized using MZI (see Fig. \ref{fig:Unified_Optical_circuit}) in which the decision to insert or remove the second beam splitter (BS$_2$) is made only after the photon has passed the first beam splitter (BS$_1$), thereby such configuration eliminates any possibility that the photon could possess prior knowledge  of the measurement based on hidden variable theory. The first experimental realizations of DC ideas emerged during the 1980s with several proposals on different physical systems including neutron interferometer \cite{rauch1984static,kawai1998realization} and photonic interferometer \cite{alley1984delayed,Mittelstaedt1986,hellmuth1987delayed,baldzuhn1989wave}. 
A significant advancement was achieved with the development of quantum-delayed choice experiment \cite{Peruzzo,tang2012,ionicioiu2011proposal,ma2016delayed}, which employed a quantum-controlled beam splitter to simultaneously investigate the wave-like and particle-like behavior of photons. 

Despite their distinct historical origins and conceptual objectives, IFM and DC experiments share several fundamental features. In IFM, the possibility of obtaining information without direct interaction modifies the interference pattern, whereas in DC experiments, the choice of measurement context determine whether wave-like or particle-like behavior is observed.  Nevertheless they are traditionally been investigated independently without seeing these paradigms from the same conceptual lens with a unified description capable of encompassing both phenomena within a single quantum architecture. Motivated by this observation, we propose a unified quantum interferometric framework that simultaneously incorporates IFM and DC behavior within a common quantum circuit representation. By extending the Elitzur-Vaidman IFM scheme with a quantum controlled DC mechanism, we demonstrate how counterfactual information extraction and wave particle duality can emerge from the same underlying interferometric structure. The theoretical predictions are validated through numerical simulations using Qiskit and further verified on IBM quantum computer and Qniverse platforms. The results provide a unified perspective on quantum measurement, complementarity, counterfactual and DC phenomena, providing new insights into the foundational principles of quantum information processing.

The rest of the paper is structured as follows: Section \ref{sec2} presents the proposed quantum unified interferometric framework, including its theoretical analysis and mathematical formulation. Section \ref{results_discussion} discusses the results obtained through numerical simulations on Qiskit, IBM quantum computer and Qniverse platforms. Section \ref{sec4} concludes the paper and outlines the possible future research directions. 

\section{Unified Framework for Interaction free measurement and  Quantum Delayed Choice}\label{sec2}

We now describe in detail the proposed unified framework for the study of IFM along with quantum DC experiment. We will first describe the theoretical framework which will then be followed by a rigorous mathematical analysis.  

\subsection{ Theoretical Framework}\label{subsec1}
\begin{figure}[!t]
    \centering
        \begin{subfigure}{\textwidth}
            \centering
            \includegraphics[width=\textwidth]{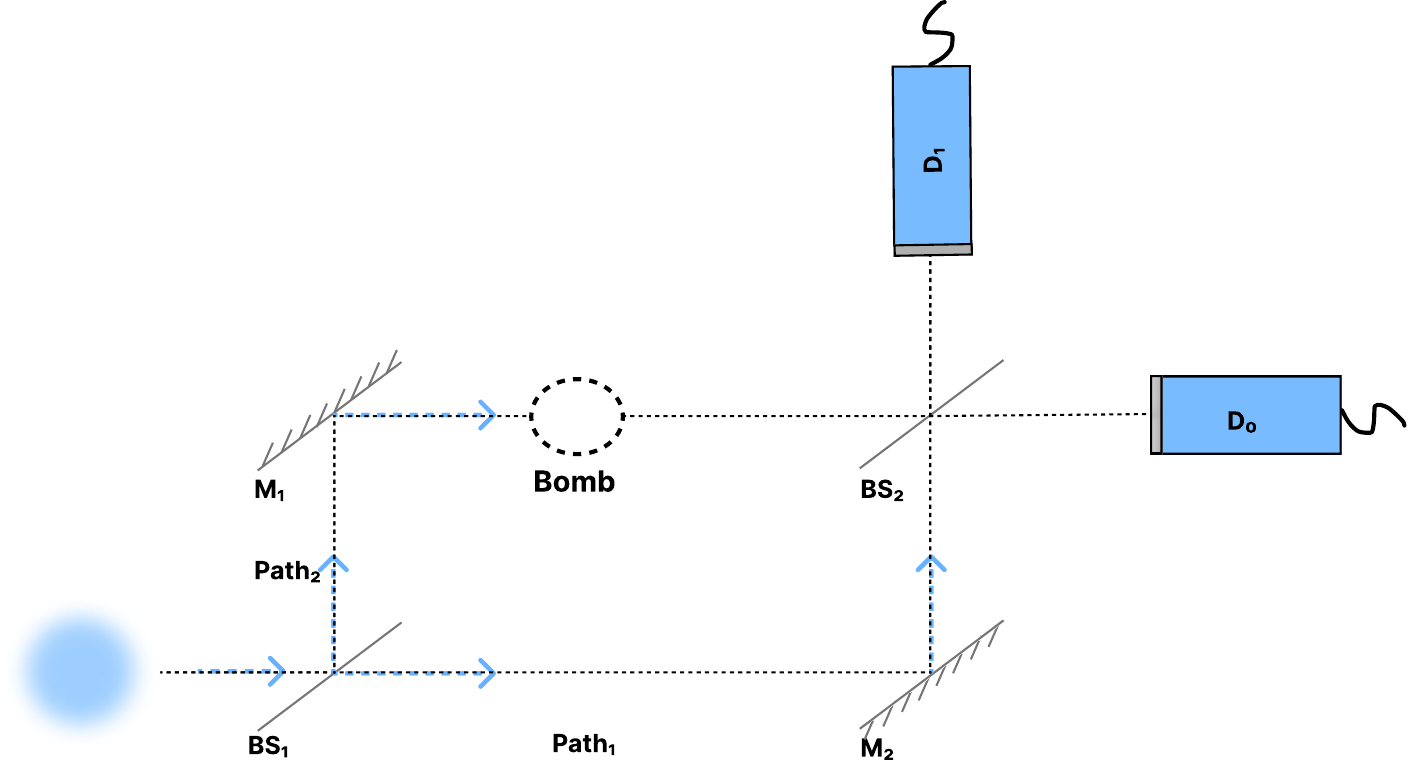}
            \caption{Optical realization of the proposed unified model based on Mach-Zehnder interferometer. A single photon is prepared in a coherent superposition of two paths by the first beam splitter BS$_1$. One arm of the interferometer contains a quantum-controlled absorber (quantum bomb), while the other arm remains unobstructed. The two paths are recombined at the second beam splitter BS$_2$, whose presence or absence is controlled in the quantum-controlled beam splitter, and the photon is subsequently detected at either detector $D_0$ or $D_1$.} \label{fig:Unified_Optical_circuit}
        \end{subfigure}
       \begin{subfigure}{\textwidth}
            \centering    
            \includegraphics[width=\textwidth]{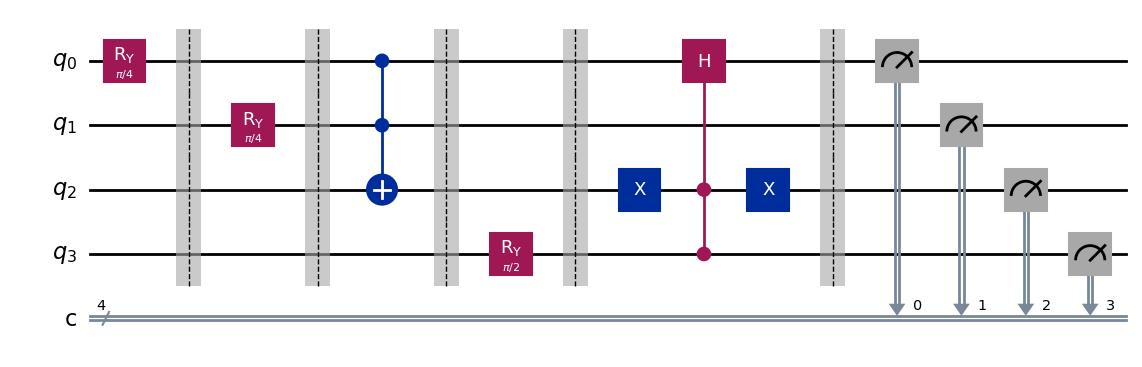}
            \caption{Equivalent quantum circuit representation of the unified model. Here, $q_0$ encodes the two optical paths of the interferometer and is parameterized by the beam splitter angle $\theta$, $q_1$ represents the state of quantum bomb and is characterized by the parameter $\beta$, $q_2$ records the explosion events and $q_3$ represents the state of quantum-controlled second beam-splitter, governed by the delayed choice parameter $\alpha$.} \label{fig:Unified_Quantum_circuit}
        \end{subfigure}
    \caption{(Color Online) Unified representation of interaction-free measurement in the delayed choice framework: (a) Optical realization and (b) Equivalent quantum circuit representation.}
    \label{fig:Unified_circuit}
\end{figure}

We develop a four qubit quantum circuit representation based on the  MZI as the underlying framework to investigate how IFM can be understood within the DC configuration. Fig.~\ref{fig:Unified_Optical_circuit} shows the optical realization of the setup, while Fig.~\ref{fig:Unified_Quantum_circuit} presents its equivalent quantum circuit representation.

In MZI, the photon is prepared in a spatial superposition using the first beam splitter  $BS_1$ (see Fig.~\ref{fig:Unified_Optical_circuit}). This is implemented via the application of single qubit rotation gate  $R_y({\theta})$ on the qubit $\ket{q_0}$ (see Fig.~\ref{fig:Unified_Quantum_circuit}). The resulting state is given by $|q_0\rangle = \cos{\theta} \ket{0} + \sin{\theta} \ket{1}$ where, $\theta \in [0,\pi/2]$ parameterizes the ``superposition of both paths created by the first beam splitter" with $\ket{0}$ representing the lower path and $\ket{1}$ representing the upper path. The state of quantum-controlled absorber (hereafter referred to as the quantum bomb)  is represented by qubit $\ket{q_1} $ which, upon application of rotation gate $R_y({\beta})$ (see Fig.~\ref{fig:Unified_Quantum_circuit}) can be expressed as $|q_{1}\rangle = \cos{\beta} \ket{0} + \sin{\beta} \ket{1}$ where $\beta \in [0,\pi/2]$  controls the relative amplitudes of the completely absent bomb state ($\ket{0}$) and completely present bomb state ($\ket{1}$). We next introduce a controlled interaction between the photon qubit $\ket{q_0}$ and bomb state qubit $\ket{q_1}$ to record whether an interaction event, i.e., photon absorption, has occurred. The qubit $\ket{q_2}$ is used to keep a record of such event. This is realized via a CCN gate (Toffoli-gate), such that $\ket{q_2}$ which is initially at $\ket{0}$ flips only when both the qubits $\ket{q_0}$ and $\ket{q_1}$ are in the state $\ket{1}$ and thereby recording the ``explosion" event (see Fig.~\ref{fig:Unified_Quantum_circuit}). 

The DC mechanism is implemented using the qubit $\ket{q_3}$ which coherently controls the insertion or removal of the second beam splitter $BS_2$ (see Fig.~\ref{fig:Unified_Optical_circuit}). The continuous transition of wave and particle like behavior in DC  is implemented via application of $R_y({\alpha})$ on $\ket{q_3}$ (see Fig.~\ref{fig:Unified_Quantum_circuit}). Further, we have to make sure that the action of $BS_2$ is restricted only to the survival events i.e., when no explosion (absorption) event has occurred. This is realized through controlled controlled Hadamard (CCH) operation on the photon qubit $\ket{q_0}$ in order to study IFM and DC experiments simultaneously. For this, we first flip the $\ket{q_2}$ ancillary qubit using the Pauli X gate, $X = \begin{pmatrix} 0 & 1 \\ 1 & 0 \end{pmatrix}$ which now temporarily labels $\ket{q_2} = \ket{1}$ as survival events and $\ket{q_2} = \ket{0}$ as explosion events, then apply the CCH gate to the photon qubit $\ket{q_0}$ where the controls are $\ket{q_2}$ and $\ket{q_3}$ qubits such that when both qubits $\ket{q_2}$ and $\ket{q_3}$ are $\ket{1}$.  Finally, the second Pauli X gate acts on $\ket{q_2}$ to restore the original physical meaning of qubit $\ket{q_2}$. The sequence (X-CCH-X) effectively realizes the action of second beam splitter (Hadamard operation) on the photon qubit $\ket{q_0}$ for  joint study of the quantum bomb under DC mechanism (see Fig.~\ref{fig:Unified_Quantum_circuit}). So, the absorbed photon will not take further participation in the unitary evolution, irrespective of whether the second beam splitter is present or absent. 

In standard DC experiment, only one ancilla qubit is used to realize the quantum control of the second beam splitter (Hadamard gate) action on photon qubit $\ket{q_0}$, which is implemented by Controlled Hadamard (CH) gate. But here, the reason to choose the CCH gate is that the bomb qubit $\ket{q_1}$ is in quantum superposition, which also affects the action of the second beam splitter on the photon qubit $\ket{q_0}$. When the bomb qubit is in present state then photon will be absorbed and the action of BS$_2$ should not be applied on the absorbed photons. The absorbed-photon branches are excluded from the detector statistics by post-selecting the surviving-photon subspace. Interference and which path information is recovered only in those post selection states in which the photon has survived. 

\subsection{ Mathematical Formulation}\label{subsec2}

Initially all the qubits are prepared in the computational basis state $\ket{\psi_0} = \ket{0000}_{q_0q_1q_2q_3}$ with first qubit \(\ket{q_0}\) representing the spatial path of photon, second qubit \(\ket{q_1}\) representing the state of quantum bomb, third qubit \(\ket{q_2}\) representing explosion event while the fourth qubit \(\ket{q_3}\) represent the state of second beam-splitter. After the application of R$_y(2\theta)$ to the photon qubit, the state can be written as 
\begin{equation}
\ket{\psi_1} = \cos\theta \ket{0000} + \sin\theta \ket{1000}.
\label{eq:BS1_action0}
\end{equation}
For the special case $\theta = \frac{\pi}{4}$, the beam splitter becomes a balanced \(50{:}50\) beam splitter, generating equal superposition between both photon paths. After the application of \(R_y(2\beta)\) on the bomb qubit, the state can be written as 
\begin{equation}    
\ket{\Psi_2} = \cos\theta \cos\beta \ket{0000} + \cos\theta \sin\beta \ket{0100} + \sin\theta \cos\beta \ket{1000} + \sin\theta \sin\beta \ket{1100}.
\label{eq:Quantum_Bomb_implementation}
\end{equation}
The interaction between the photon and the bomb is modeled using a CCX gate. The photon qubit \(\ket{q_0}\) and bomb qubit \(\ket{q_1}\) act as control qubits, while the ancilla qubit \(\ket{q_2}\) acts as the target. Therefore, the resulting state can be written as 
\begin{equation}
\ket{\Psi_3} = \cos\theta \cos\beta \ket{0000} + \cos\theta \sin\beta \ket{0100} + \sin\theta \cos\beta \ket{1000} + \sin\theta \sin\beta \ket{1110}.
\label{eq:Quantum_Bomb_absorption}
\end{equation}
Unlike the conventional Elitzur--Vaidman model involving irreversible photon absorption, the present framework encodes the interaction coherently through ancilla entanglement. Consequently, the overall circuit evolution remains unitary. The presence and absence of second beam-splitter is implemented through the application of \(R_y(2\alpha)\) on qubit  \(\ket{q_3}\). To implement the quantum-controlled DC mechanism, the interaction ancilla qubit \(\ket{q_2}\) is temporarily inverted using a Pauli-\(X\) gate. This operation changes the control condition of the subsequent CCH operation such that interference is applied only to the non-interaction branch of the evolution. After the application of this operation, the state can be written as  
\begin{equation}
\begin{aligned}
\ket{\Psi_4}
=
&
\,\cos\alpha
\Big(
\cos\theta\cos\beta \ket{0010}
+
\cos\theta\sin\beta \ket{0110}
\\
&
+
\sin\theta\cos\beta \ket{1010}
+
\sin\theta\sin\beta \ket{1100}
\Big)
\\
&
+
\sin\alpha
\Big(
\cos\theta\cos\beta \ket{0011}
+
\cos\theta\sin\beta \ket{0111}
\\
&
+
\sin\theta\cos\beta \ket{1011}
+
\sin\theta\sin\beta \ket{1101}
\Big).
\end{aligned}
\label{eq:Delayed_choice_implementation_b}
\end{equation}
CCH gate is then applied to the photon qubit \(\ket{q_0}\).  Physically, this operation simulates the conditional insertion or removal of the second beam splitter BS\(_2\). Finally, the interaction ancilla \(\ket{q_2}\) is restored to its original encoding using another Pauli-\(X\) operation. The final quantum state of the complete interferometric system can be written as 
\begin{equation}
\begin{aligned}
\ket{\Psi_{\mathrm{final}}}
=
&
\,\cos\alpha
\Big(
\cos\theta\cos\beta \ket{0000}
+
\cos\theta\sin\beta \ket{0100}
\\
&
+
\sin\theta\cos\beta \ket{1000}
+
\sin\theta\sin\beta \ket{1110}
\Big)
\\
&
+
\frac{\sin\alpha}{\sqrt{2}}
\Big[
(\cos\theta+\sin\theta)\cos\beta \ket{0001}
\\
&
+
(\cos\theta-\sin\theta)\cos\beta \ket{1001}
\\
&
+
\cos\theta\sin\beta
\left(
\ket{0101}
+
\ket{1101}
\right)
\Big]
\\
&
+
\sin\alpha\sin\theta\sin\beta
\ket{1111}.
\end{aligned}
\label{eq:Delayed_choice_implementation_c}
\end{equation}
The final state contains all physically relevant outcomes of the unified framework which includes IFM branches, interference and non-interference evolution, quantum-controlled DC behavior, and photon-bomb interaction (explosion) events.

To obtain a physically meaningful detector statistics, post-selection is performed by considering only the survival branches of the quantum state corresponding to photons that successfully propagate to the detectors. Interaction or explosion branches are excluded from the normalization because those events terminate the interferometric evolution and therefore do not contribute to observable detector clicks. The post-selected survival state can be written as
\begin{equation}
\begin{aligned}
\ket{\Psi_{\mathrm{surv}}}
=
&
\,\cos\alpha
\Big(
\cos\theta\cos\beta \ket{0000}
+
\cos\theta\sin\beta \ket{0100}
\\
&
+
\sin\theta\cos\beta \ket{1000}
\Big)
\\
&
+
\frac{\sin\alpha}{\sqrt{2}}
\Big[
\cos\beta(\cos\theta+\sin\theta)\ket{0001}
\\
&
+
\cos\beta(\cos\theta-\sin\theta)\ket{1001}
\\
&
+
\cos\theta\sin\beta
\left(
\ket{0101}
+
\ket{1101}
\right)
\Big].
\end{aligned}
\label{eq:Post_selection}
\end{equation}
Since, $\theta$ and $\beta$ are the parameters that respectively control the two paths of MZI and the state of quantum bomb, so the probability associated with the absorption of photon by the quantum bomb can be written as
\begin{equation}
P_{\mathrm{abs}} = \sin^2\theta \sin^2\beta.   
\end{equation}
Accordingly, the probability for the survival of the photon can be written as
\begin{equation}
\begin{aligned}
P_{\mathrm{surv}}
&=
1 - \sin^2\theta \sin^2\beta
\\
&=
\braket{\Psi_{\mathrm{surv}}|\Psi_{\mathrm{surv}}}.
\end{aligned}
\end{equation}
Now, the post-selected quantum state (Eq. \ref{eq:Post_selection}) can be normalized and the normalized post-selected quantum state can be written as 
\begin{equation}
\ket{\Psi_{\mathrm{post}}}
=
\frac{
\ket{\Psi_{\mathrm{surv}}}
}{
\sqrt{
1-\sin^2\theta\sin^2\beta
}
}.
\label{eq:Postselection_state}
\end{equation}
This normalized post-selected state represents the conditional interferometric evolution of photons that successfully survive and reach the output detectors.

\begin{table}[h!]
\centering
\begin{tabular}{|c|c|c|}
\hline
\textbf{State} & \textbf{Probability} & \textbf{Physical Interpretation} \\
\hline
$\ket{0000}$ & 0.125 & Superposition collapse $D_0$ \\
\hline
$\ket{0100}$ & 0.125 & Photon survive wihout bomb interaction \\
\hline
$\ket{1000}$ & 0.125 &  Superposition collapse $D_1$ \\
\hline
$\ket{1110}$ & 0.125 & Explosion event recorded by ancilla \\
\hline
$\ket{0001}$ & 0.25 & Constructive interference outcome \\
\hline
$\ket{1001}$ & 0 & Destructive interference outcome \\
\hline
$\ket{0101}$ & 0.0625 & Inconclusive interaction-free branch \\
\hline
$\ket{1101}$ & 0.0625 & Conclusive interaction-free detection outcome \\
\hline
$\ket{1111}$ & 0.125 & Explosion event branch \\
\hline
\end{tabular}
\caption{Detector statistics for both BS$_2$ present ($\alpha=\pi/2$) and absent ($\alpha = 0$) cases.}
\label{tab1}
\end{table}
To further verify the universality of the proposed model, the quantum circuit was executed on Qiskit \textit{Simulator} \cite{qiskit2026}, and IBM Quantum hardware \cite{IBMQuantum2026}. Further, the circuit was translated into a bulk-optics photonic architecture using the Qniverse  \textit{Photonic simulator} \cite{Qniverse2026}.  Unlike the IBM Quantum hardware execution, the Qniverse simulation operates in an idealized environment without introducing the gate noise and hardware induced decoherence. 
\begin{figure}[H]
    \centering
    \includegraphics[width=\textwidth]{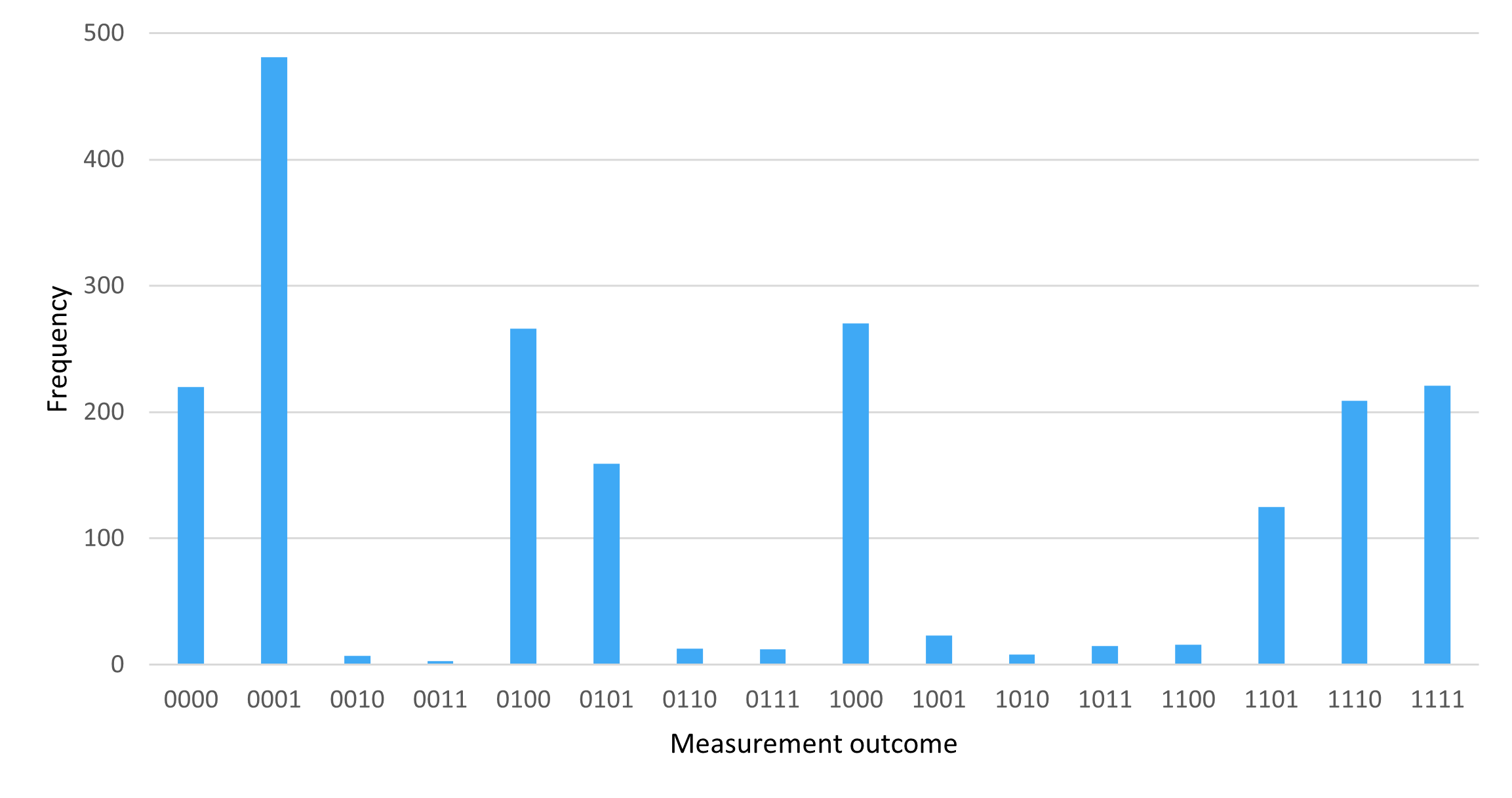}
    \caption{Detector counts obtained from execution on the IBM Quantum hardware.} 
    \label{fig:ibm_backend}
\end{figure}
\begin{figure}[H]
    \centering
    \includegraphics[width=0.75\textwidth]{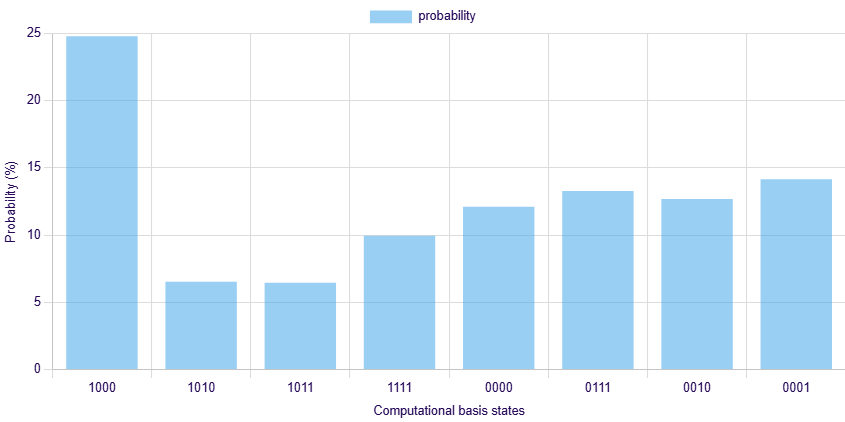}
    \caption{Detector probability distribution obtained using the Qniverse platform simulator.}
    \label{fig:qniverse_setup}
\end{figure}

\section{Results and Discussion} \label{results_discussion}

The theoretical structure proposed and analyzed in Section \ref{sec2} provides a complete description of the proposed unified interferometric framework, enabling the evaluation of detector probabilities. In this section, the analytical predictions are validated through numerical simulations using the Qiskit statevector simulator \cite{qiskit2026}, IBM Quantum hardware \cite{IBMQuantum2026}, and Qninverse simulator \cite{Qniverse2026}. The detector probabilities are analyzed as a function of the circuit parameters ($\theta,\beta,\alpha$), demonstrating the continuous transition between particle-like and wave-like behaviors and the emergence of IFM within the unified interferometric framework.

\begin{figure}[htbp]
\centering
    \begin{subfigure}{0.7\columnwidth}
        \centering
        \includegraphics[width=0.7\linewidth]{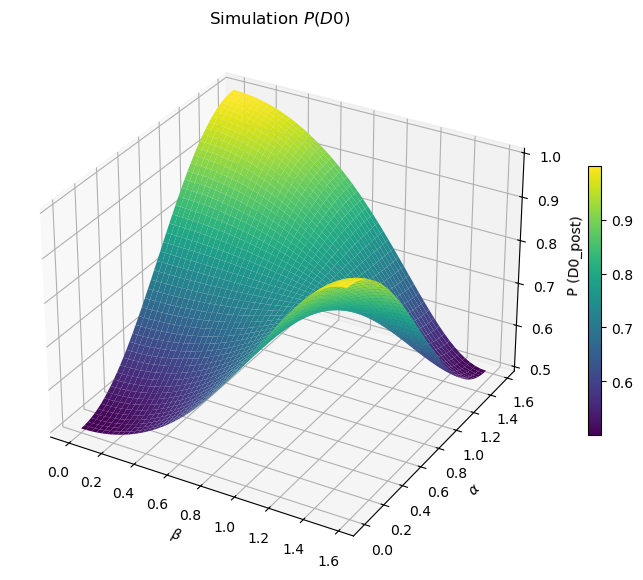}
        \caption*{(a)}
        \label{fig:d0_simulation}
    \end{subfigure}
   \vspace{1em}
    \begin{subfigure}{0.7\columnwidth}
        \centering
        \includegraphics[width=0.7\linewidth]{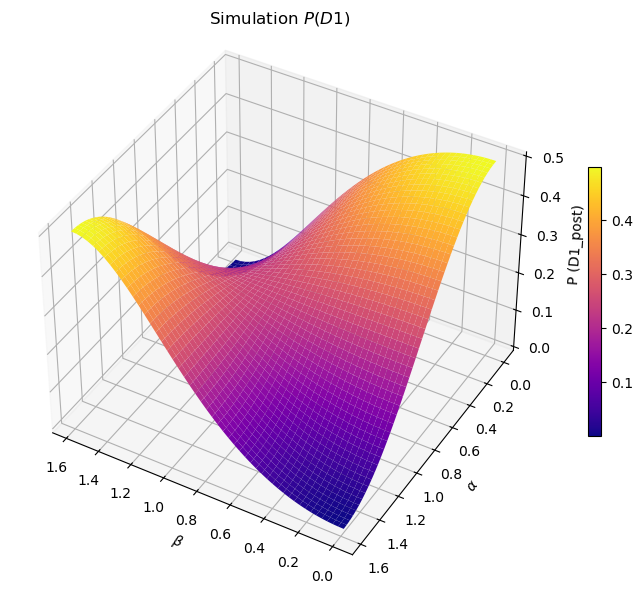}
        \caption*{(b)}

        \label{fig:d1_simulation}
    \end{subfigure}    
    \caption{(Color online) The detector probabilities (a) \(P(D_0)\) and (b) \(P(D_1)\) for the DC wave-particle transition in the unified  interferometric configuration \((\theta=\pi/4)\), plotted as functions of the DC parameter \(\alpha\) and quantum bomb state parameter \(\beta\). The results demonstrate the gradual transition from particle-like behavior to wave-like interference as the second beam splitter BS$_2$ is coherently present. The numerical simulations were performed using Qiskit's ideal statevector simulation which implements noiseless unitary evolution. Consequently, the simulated detector probabilities agree with the analytical predictions.}    
    \label{fig:wave_particle_transition_comparison}
\end{figure}

 To verify the analytical predictions derived in Section \ref{sec2}, the proposed quantum circuit was implemented using three computational platforms: the Qiskit statevector simulator (see Table. \ref{tab1}), IBM Quantum hardware (see Fig. \ref{fig:ibm_backend}) and the Qniverse simulator (see Fig. \ref{fig:qniverse_setup}). These implementations provide independent verifications of the theoretical framework while allowing the influence of realistic hardware imperfection to be distinguished from the ideal unitary evolution. The detector probabilities obtained from the simulations reproduce those predicted by the analytical expressions. In particular, the simulated detector probabilities shown in Fig.~\ref{fig:wave_particle_transition_comparison} agree with the analytical predictions over the entire range of the DC parameter $\alpha$ and the quantum bomb state parameter $\beta$.
To investigate the practical feasibility of the proposed framework the same circuit was executed on IBM Quantum hardware and measured counts are presented in Fig. \ref{fig:ibm_backend}. Despite the experimental imperfections (gate errors, qubit decoherence, readout noise), the principal features predicted by the analytical model remain clearly observable. The detector probabilities continue to exhibit the expected transition between particle-like and wave-like behaviour, while the interaction-free measurement branches remain distinguishable from the explosion events. The preservation of these characteristic signatures indicates that the proposed framework can be implemented on present-day noisy intermediate-scale quantum (NISQ) devices, although a systematic noise and scalability analysis is required to establish its robustness. The validation of the results was also done using Qniverse simulator. This simulator does not introduce any gate errors or noise. The probability distribution of the obtained measurement outcomes is presented in Fig. \ref{fig:qniverse_setup}. We can clearly see that the simulated probability distributions agree almost exactly with the analytical detector statistics summarized in Table \ref{tab1}. 

The present framework represents the second beam splitter by a quantum-controlled operation, allowing the interferometer to evolve continuously between the open and closed configuration which is governed by the parameter ($\alpha$). The detector probabilities obtained from the post selection state Eq~\eqref{eq:Postselection_state} can be written as,
\begin{equation}
    P_{post}(D_0)
=\frac{\cos^2\alpha \cos^2\theta + \frac{1}{2}\sin^2\alpha\big[\cos^2\beta[1+\sin2\theta] + \cos^2\theta \sin^2\beta \big]}{1-\sin^2\theta\sin^2\beta},
\end{equation}
and 
\begin{equation}
P_{post}(D_1)
=
\frac{\cos^2\alpha \sin^2\theta \cos^2\beta
+
\frac{1}{2}\sin^2\alpha\big[\cos^2\beta[1-\sin2\theta] + \cos^2\theta \sin^2\beta \big]}{1-\sin^2\theta\sin^2\beta}.
\end{equation}
These expressions reveal that the output statistics depend continuously on the DC parameter.
For the limiting case ($\alpha=0$), the second beam splitter BS$_2$ is effectively absent, and the interferometer behaves as an open Mach-Zehnder interferometer. In this regime, the photon retains complete which-path information and detector probabilities reduces to 
\begin{equation}
P(D_0)= \frac{\cos^2\theta}{1-\sin^2\theta\sin^2\beta}, \quad P(D_1) = \frac{\sin^2\theta\cos^2\beta}{1-\sin^2\theta\sin^2\beta}.
\end{equation}
For balanced configuration $(\theta = \pi/4, \beta = 0)$, these simplify to $P(D_0)=P(D_1) = \frac{1}{2}$, indicating equal detection probabilities and the absence of interference, which is characteristic of particle-like behavior.

As $\alpha$ increases, the second beam splitter is introduced coherently allowing the probability amplitudes associated with the two paths to interfere. Consequently, the detector probabilities evolve continuously from the open to closed interferometer regime, illustrating the gradual transition between particle-like and wave-like behavior governed by the quantum-controlled measurement configuration. For other opposite limiting case, $\alpha=\frac{\pi}{2}$, the interferometer is closed and detector probabilities reduces to
\begin{equation}
 P(D_0)=\frac12 \Bigg[\frac{\cos^2\beta(1+\sin2\theta) + \cos^2\theta\sin^2\beta}{1-\sin^2\theta\sin^2\beta}\Bigg],   
\end{equation}
and
\begin{equation}
P(D_1)=\frac12 \Bigg[\frac{\cos^2\beta(1-\sin2\theta) + \cos^2\theta\sin^2\beta}{1-\sin^2\theta\sin^2\beta}\Bigg].   
\end{equation}
For balanced configuration $(\theta = \pi/4, \beta = 0)$, these reduce to $P(D_0)= 1$ and $P(D_1) = 0$, demonstrating complete constructive and complete destructive interference at the detectors $D_0$ and $D_1$, respectively, which is characteristic of wave-like behavior. The continuous evolution of the detector probabilities is illustrated in Fig.~\ref{fig:wave_particle_transition_comparison}, which shows the dependence of the detector probabilities on the DC parameter $\alpha$.

The quantum bomb state parameter $\beta$ determines the state of quantum bomb. Unlike the original Elitzur-Vaidman bomb tester, the proposed framework allows bomb to be in quantum superposition state of present and absent state. Consequently, the interferometer continuously undergo transition between the interference and IFM. This parameter provides a direct measure of the probability that the photon encounters an absorbing object while propagating through one arm. When $\beta=0$ (bomb is complete absent), the photon propagates coherently through both interferometer arms, and the interference pattern depends solely on the DC parameter $\alpha$. In particular, for $\alpha=\pi/2$ the interferometer reproduces the results of standard closed Mach-Zehnder interferometer i.e., complete constructive interference at detector $D_0$ and complete destructive interference at detector $D_1$. 

As the parameter $\beta$ increases, the probability of photon-bomb interaction also increases. The availability of the path information reduces the coherence between the two interferometer arms, resulting in a gradual suppression of the interference pattern. The loss of interference therefore arises from the photon-bomb interaction events. Whenever photon propagates through the bomb arm, and interaction may occur and the explosion events is capture by the ancilla $\ket{q_2}$. After post selection, only the surviving photon branches contribute to the observed detector statistics. The resulting detector probabilities therefore no longer exhibit complete interference, despite the presence of the second beam splitter. The numerical probability distribution shown in Fig.~\ref{fig:wave_particle_transition_comparison} clearly demonstrates this behavior. As bomb parameter $\beta$ increases, the interference pattern gradually diminishes until the characteristics of IFM regime are reached. The smooth variation of the detector probabilities confirms that the proposed framework provides a continuous transition between conventional interferometer and counterfactual measurement rather than treating them as two independent experimental configuration.

A distinguishable feature of the proposed unified framework is its ability to capture the IFM while simultaneously incorporating the DC features. Unlike the conventional Elitzur-Vaidman bomb testing experiment, where the absorbing object is treated as a classical obstacle, the present framework allows the bomb to exist in a coherent quantum superposition. Consequently, the probability of IFM can be continuously controlled through the quantum bomb state parameter $\beta$, while the manifestation of interference is independently governed by the DC parameter $\alpha$.
Following post-selection procedure described in Section \ref{sec2}, only those branches corresponding to surviving photons are retained for the evaluations of detector statistics. The conditional IFM probability can be calculated as
\begin{equation}
 P(IFM|survival) =  \frac{\sin^2\alpha\cos^2\theta\sin^2\beta}{2(1-\sin^2\theta\sin^2\beta)}.   
\end{equation}
This expression explicitly demonstrates that the IFM probability depends on both the DC parameter and quantum bomb state parameter.

\begin{figure}[htbp]
    \centering
    \includegraphics[width=0.7\textwidth]{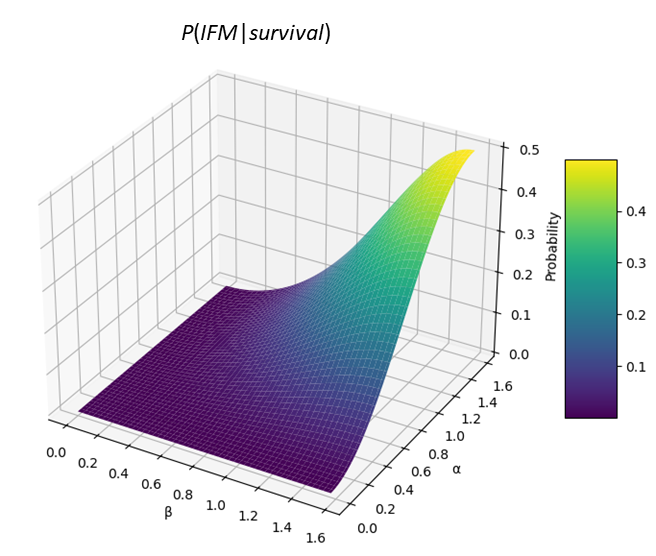}
    \caption{(Color online) Post-selected conditional IFM probability $P(IFM|survival)$ as a function of the DC parameter $\alpha$ and bomb state parameter $\beta$ for the balanced interferometric configuration $\theta=\pi/4$.
    The probability varies smoothly with both control parameters and attains its maximum value of $\frac{1}{2}$ for the configuration $\alpha = \beta = \pi/2$, corresponding to the conventional IFM regime conditioned on survived photons.}
    \label{fig:ifm_surface}
\end{figure}
To illustrate the physical significance of this results, several limiting cases may be considered.
For $\beta=0$, the bomb is completely absent from the interferometer. Since, no absorbing object exists, no IFM can occur, and the conditional IFM vanishes i.e. $P(IFM|survival) =  0$. Under these conditions, the interferometer behaves identically to the conventional DC experiment with the detector statistics determined solely by the parameter $\alpha$. As the value of the $\beta$ increases as shown in Fig. \ref{fig:ifm_surface}, the quantum bomb state gradually influences the interferometric evolution. The probability of photon interacting with the bomb correspondingly increases, and the probability of obtaining which path information also increases. When $\beta=\pi/2$ and $\alpha=\pi/2$, the system is in IFM configuration and for 50:50 BS i.e., $\theta=\pi/4$ the conditional IFM probability becomes equal to $\frac{1}{2}$. This value reflects the probability of obtaining an IFM conditioned on photon survival after post-selection. The corresponding unconditional IFM probability is $P(IFM) = P(IFM|survival) \cdot P(survival) = \frac{1}{2} \cdot \frac{1}{2} =\frac{1}{4}$ thereby account for the characteristics $25 \%$ IFM detection efficiency predicted by the original EV bomb testing experiment.

\section{Conclusion}\label{sec4}
In this work, we have developed a unified quantum interferometric framework that integrates IFM and the quantum DC experiment. A four-qubit quantum circuit formulation is developed, leading to analytical expressions for the detector probabilities, survival probability, absorption probability, and the post-selected IFM probability. This model provides a continuous description of the transition between wave-like and particle-like behavior while simultaneously incorporating interaction-free measurement through the control of the bomb-state parameter $\beta$. The proposed framework offers a unified physical interpretation of complementarity, which-path information, and counterfactual measurement. The DC parameter $\alpha$ governs the coherent transition between interference and path distinguishability. The unified description highlights that the wave-particle duality and counterfactual information extraction can be manifested simultaneously from the same quantum evolution governed by the conditional quantum operations. Consequently, the present work provides a common theoretical perspective of IFM and DC in interferometric setup.

The analytical predictions are validated through numerical simulations using the Qiskit statevector simulator, IBM quantum hardware and  Qniverse photonic simulator. The simulations using Qiskit and Qniverse quantum simulator exhibit excellent agreement with the analytical model,providing numerical support for the analytical predictions of the proposed model. As expected, quantum circuit implementation using the IBM quantum hardware shows some  small deviations arising due to gate errors, readout noise and decoherence characteristics of noisy intermediate scale quantum (NISQ) devices. Beyond its foundational significance, the proposed circuit-based formulation establishes a foundation for investigating broader applications of IFM in quantum technology such as counterfactual quantum communication, quantum key distribution, quantum secure direct communication and quantum zeno effect. Further, this model has the potential of experimental realization using bulk-optics, thus extending the model to include realistic decoherence, optical losses, imperfect detectors. Such investigations will enable a more comprehensive assessment of the robustness and scalability of the framework under practical experimental conditions and may further advance its applications in quantum technologies.

\section*{Acknowledgment}
The authors thank the Department of Science and Technology, Government of India, for the support provided through the National Quantum Mission (NQM). The authors thank Prof. Tabish Qureshi for useful discussions.

\section*{Competing Interests}
Authors declare that they don't have any competing interests.

\section*{Data Availability}
No additional data is generated through this work. All the relevant data is included in the paper.

\bibliography{main}%

@article{debroglie1925,
  author     = "de Broglie, Louis",
  title      = "Recherches sur la théorie des quanta",
  journal    = "Annales de Physique",
  volume     = "10",
  number     = "3", 
  pages      = "22--128",
  year       = "1925",
  doi        = "10.1051/anphys/192510030022",
  note       = "Thèse de doctorat, Université de Paris, 1924"
}

@article{Bohr1928,
  author    = "Bohr, N.",
  title     = "The Quantum Postulate and the Recent Development of Atomic Theory",
  journal   = "Nature",
  volume    = "121",
  number    = "3050",
  pages     = "580--590",
  year      = "1928",
  doi       = "10.1038/121580a0"
}

@article{Heisenberg1927,
  author    = "Heisenberg, W.",
  title     = "{\"{U}}ber den anschaulichen {I}nhalt der quantentheoretischen {K}inematik und {M}echanik",
  journal   = "Zeitschrift f{\"u}r Physik",
  volume    = "43",
  number    = "3",
  pages     = "172--198",
  year      = "1927",
  doi       = "10.1007/BF01397280"
}

@article{Scully,
  title = {Quantum eraser: A proposed photon correlation experiment concerning observation and ``delayed choice" in quantum mechanics},
  author = {Scully, Marlan O. and Dr\"uhl, Kai},
  journal = "Physical Review A",
  volume = "25",
  issue = "4",
  pages = "2208--2213",
  year = "1982",
  month = "Apr",
  publisher = "American Physical Society",
  doi = "10.1103/PhysRevA.25.2208"
}

@article{elitzur1993quantum,
  title="Quantum mechanical interaction-free measurements",
  author="Elitzur, Avshalom C. and Vaidman, Lev",
  journal="Found Phys",
  volume="23",
  number="7",
  pages="987--997",
  year="1993",
  publisher="Springer",
  doi="10.1007/BF00736012"
}

@article{Kwiat,
  title = {High-Efficiency Quantum Interrogation Measurements via the Quantum Zeno Effect},
  author = {Kwiat, P. G. and White, A. G. and Mitchell, J. R. and Nairz, O. and Weihs, G. and Weinfurter, H. and Zeilinger, A.},
  journal = {Physical Review Letters},
  volume = {83},
  issue = {23},
  pages = {4725--4728},
  numpages = {0},
  year = {1999},
  month = {Dec},
  publisher = {American Physical Society},
  doi = {10.1103/PhysRevLett.83.4725},
  url = {https://link.aps.org/doi/10.1103/PhysRevLett.83.4725}
}

@article{zeno,
  author  = {Misra, B. and Sudarshan, E. C. G.},
  title   = {The Zeno's paradox in quantum theory},
  journal = {Journal of Mathematical Physics},
  year    = {1977},
  volume  = {18},
  number  = {4},
  pages   = {756--763},
  doi     = {10.1063/1.523304},
  url     = {https://doi.org/10.1063/1.523304},
  publisher = {AIP Publishing},
  month   = {April}
}

@misc{bennett1984,
  author = "Bennett, C. H. and Brassard, G.",
  title  = "Quantum Cryptography: Public Key Distribution and Coin Tossing",
  year   = "1984",
  note   = "Paper presented at the IEEE International Conference on Computers, Systems and Signal Processing, Bangalore, India,(IEEE, New York 1984), 10--12 December 1984, pp. 175--179"
}

@article{bennett1992,
  title = {Quantum cryptography without {B}ell's theorem},
  author = {Bennett, Charles H. and Brassard, Gilles and Mermin, N. David},
  journal = {Physical Review Letters},
  volume = {68},
  issue = {5},
  pages = {557--559},
  numpages = {0},
  year = {1992},
  month = {Feb},
  publisher = {American Physical Society},
  doi = {10.1103/PhysRevLett.68.557},
}

@article{E91,
  title = {Quantum cryptography based on {B}ell's theorem},
  author = {Ekert, Artur K.},
  journal = {Physical Review Letters},
  volume = {67},
  issue = {6},
  pages = {661--663},
  numpages = {0},
  year = {1991},
  month = {Aug},
  publisher = {American Physical Society},
  doi = {10.1103/PhysRevLett.67.661},
  url = {https://link.aps.org/doi/10.1103/PhysRevLett.67.661}
}

@article{GV,
  title = {Quantum Cryptography Based on Orthogonal States},
  author = {Goldenberg, Lior and Vaidman, Lev},
  journal = {Physical Review Letters},
  volume = {75},
  issue = {7},
  pages = {1239--1243},
  numpages = {0},
  year = {1995},
  month = {Aug},
  publisher = {American Physical Society},
  doi = {10.1103/PhysRevLett.75.1239},
  url = {https://link.aps.org/doi/10.1103/PhysRevLett.75.1239}
}

@article{GUOSHi,
author = "Guang-Can Guo and Bao-Sen Shi",
title = "Quantum cryptography based on interaction-free measurement",
journal = "Physics Letters A",
volume = "256",
number = "2",
pages = "109-112",
year = "1999",
issn = "0375-9601",
doi = "https://doi.org/10.1016/S0375-9601(99)00235-2",
}

@article{NOH,
  title = "Counterfactual Quantum Cryptography",
  author = "Noh, Tae-Gon",
  journal = "Physical Review Letters",
  volume = "103",
  issue = "23",
  pages = "230501",
  numpages = "4",
  year = "2009",
  month = "Dec",
  publisher = "American Physical Society",
  doi = "10.1103/PhysRevLett.103.230501",
}

@article{ren2011experimental,
  author  = {Ren, M. and Wu, G. and Wu, E. and others},
  title   = {Experimental demonstration of counterfactual quantum key distribution},
  journal = {Laser Physics},
  year    = {2011},
  volume  = {21},
  pages   = {755--760},
  doi     = {10.1134/S1054660X11070267}
}

@article{Brida_2012,
author = {Brida, G and Cavanna, A and Degiovanni, I P and Genovese, M and Traina, P},
title = {Experimental realization of counterfactual quantum cryptography},
journal = {Laser Physics Letters},
year = {2012},
month = {jan},
volume = {9},
number = {3},
pages = {247},
doi = {10.1002/lapl.201110120},
}

@article{Salih,
  author = {Salih, Hatim and Li, Zheng-Hong and Al-Amri, M. and Zubairy, M. Suhail},
  journal = {Physical Review Letters},
  title = {Protocol for Direct Counterfactual Quantum Communication},
  volume = {110},
  issue = {17},
  pages = {170502},
  numpages = {5},
  year = {2013},
  month = {Apr},
  publisher = {American Physical Society},
  doi = {10.1103/PhysRevLett.110.170502},
  url = {https://link.aps.org/doi/10.1103/PhysRevLett.110.170502}
}

@incollection{WHEELER19789,
author  = "Wheeler, John Archibald",
title = "The “Past” and the “Delayed-Choice” Double-Slit Experiment",
editor = "A.R. Marlow",
booktitle = "Mathematical Foundations of Quantum Theory",
pages = "9-48",
address    = "New York",
publisher = "Academic Press",
year = "1978",
isbn = "978-0-12-473250-6",
doi = "https://doi.org/10.1016/B978-0-12-473250-6.50006-6",
}

@inproceedings{alley1984delayed,
  title={A delayed random choice quantum mechanics experiment with light quanta},
  author={Alley, CO and Jakubowicz, O and Steggerda, CA and Wickes, WC},
  booktitle={Proceedings of the international symposium foundations of quantum mechanics in the light of new technology},
  pages={158--164},
  year={1984}
}

@inproceedings{Mittelstaedt1986,
   author    = {Peter Mittelstaedt},
  title     = {Delayed-Choice Experiments and the Logical Analysis of Quantum Mechanics},
  booktitle = {Foundations of Quantum Mechanics in the Light of New Technology:
               Proceedings of the 2nd International Symposium},
   pages     = {53--58},
  year      = {1987},
  publisher = {Physical Society of Japan},
  address   = {Tokyo, Japan},
  }

@article{Peruzzo,
author = {Alberto Peruzzo  and Peter Shadbolt  and Nicolas Brunner  and Sandu Popescu  and Jeremy L. O’Brien },
title = {A Quantum Delayed-Choice Experiment},
journal = {Science},
volume = {338},
number = {6107},
pages = {634-637},
year = {2012},
doi = {10.1126/science.1226719},
}

@article{tang2012,
  title={Realization of quantum Wheeler's delayed-choice experiment},
  author={Tang, Jian-Shun and Li, Yu-Long and Xu, Xiao-Ye and Xiang, Guo-Yong and Li, Chuan-Feng and Guo, Guang-Can},
  journal={Nature Photonics},
  volume={6},
  number={9},
  pages={600--604},
  year={2012},
  publisher={Nature Publishing Group UK London},
  doi     = {https://doi.org/10.1038/nphoton.2012.179}
}

@article{ionicioiu2011proposal,
  title={Proposal for a quantum delayed-choice experiment},
  author={Ionicioiu, Radu and Terno, Daniel R},
  journal={Physical Review Letters},
  volume={107},
  number={23},
  pages={230406},
  year={2011},
  publisher={APS},
   doi     = {https://doi.org/10.1103/PhysRevLett.107.230406}
}

@article{pirandola2020advances,
  title={Advances in quantum cryptography},
  author={Pirandola, Stefano and Andersen, Ulrik L and Banchi, Leonardo and Berta, Mario and Bunandar, Darius and Colbeck, Roger and Englund, Dirk and Gehring, Tobias and Lupo, Cosmo and Ottaviani, Carlo and others},
  journal={Advances in optics and photonics},
  volume={12},
  number={4},
  pages={1012--1236},
  year={2020},
  publisher={Optical Society of America},
  doi= "https://doi.org/10.1364/AOP.361502",  
}

@article{rao2024protocols,
  title={Protocols for counterfactual and twin-field quantum digital signatures},
  author={Rao, Vinod N and Utagi, Shrikant and Pathak, Anirban and Srikanth, R},
  journal={Physical Review A},
  volume={109},
  number={3},
  pages={032435},
  year={2024},
  publisher={APS},
  doi="https://doi.org/10.1103/PhysRevA.109.032435",
}

@article{vaidman2019analysis,
  title={Analysis of counterfactuality of counterfactual communication protocols},
  author={Vaidman, Lev},
  journal={Physical Review A},
  volume={99},
  number={5},
  pages={052127},
  year={2019},
  publisher={APS},
  doi="https://doi.org/10.1103/PhysRevA.99.052127",
}

@article{rauch1984static,
  title={Static versus time-dependent absorption in neutron interferometry},
  author={Rauch, H and Summhammer, J},
  journal={Physics Letters A},
  volume={104},
  number={1},
  pages={44--46},
  year={1984},
  publisher={Elsevier},
  doi="https://doi.org/10.1016/0375-9601(84)90586-3",
}

@article{hellmuth1987delayed,
  title={Delayed-choice experiments in quantum interference},
  author={Hellmuth, Thomas and Walther, Herbert and Zajonc, Arthur and Schleich, Wolfgang},
  journal={Physical Review A},
  volume={35},
  number={6},
  pages={2532},
  year={1987},
  publisher={APS},
    doi="https://doi.org/10.1103/PhysRevA.35.2532",
}

@article{baldzuhn1989wave,
  title={A wave-particle delayed-choice experiment with a single-photon state},
  author={Baldzuhn, J and Mohler, E and Martienssen, W},
  journal={Zeitschrift f{\"u}r Physik B Condensed Matter},
  volume={77},
  number={2},
  pages={347--352},
  year={1989},
  publisher={Springer},
   doi="https://doi.org/10.1007/BF01313681",
}

@article{ma2016delayed,
  title={Delayed-choice gedanken experiments and their realizations},
  author={Ma, Xiao-song and Kofler, Johannes and Zeilinger, Anton},
  journal={Reviews of Modern Physics},
  volume={88},
  number={1},
  pages={015005},
  year={2016},
  publisher={APS},
   doi="https://doi.org/10.1103/RevModPhys.88.015005",
}

@article{kawai1998realization,
  title={Realization of a delayed choice experiment using a multilayer cold neutron pulser},
  author={Kawai, Takeshi and Ebisawa, Toru and Tasaki, Seiji and Hino, Masahiro and Yamazaki, Dai and Akiyoshi, Tsunekazu and Matsumoto, Yoko and Achiwa, Norio and Otake, Yoshie},
  journal={Nuclear Instruments and Methods in Physics Research Section A: Accelerators, Spectrometers, Detectors and Associated Equipment},
  volume={410},
  number={2},
  pages={259--263},
  year={1998},
  publisher={Elsevier},
  doi="https://doi.org/10.1016/S0168-9002(98)00263-0",
}

@article{englert1996fringe,
  title={Fringe visibility and which-way information: An inequality},
  author={Englert, Berthold-Georg},
  journal={Physical Review Letters},
  volume={77},
  number={11},
  pages={2154},
  year={1996},
  publisher={APS},
   doi="https://doi.org/10.1103/PhysRevLett.77.2154",
}

@article{ma2013quantum,
  title={Quantum erasure with causally disconnected choice},
  author={Ma, Xiao-Song and Kofler, Johannes and Qarry, Angie and Tetik, Nuray and Scheidl, Thomas and Ursin, Rupert and Ramelow, Sven and Herbst, Thomas and Ratschbacher, Lothar and Fedrizzi, Alessandro and others},
  journal={Proceedings of the National Academy of Sciences},
  volume={110},
  number={4},
  pages={1221--1226},
  year={2013},
  publisher={National Academy of Sciences},
  doi="https://doi.org/10.1073/pnas.1213201110",
}

@article{kwiat1995interaction,
  title={Interaction-free measurement},
  author={Kwiat, Paul and Weinfurter, Harald and Herzog, Thomas and Zeilinger, Anton and Kasevich, Mark A},
  journal={Physical Review Letters},
  volume={74},
  number={24},
  pages={4763},
  year={1995},
  publisher={APS},
   doi=" https://doi.org/10.1103/PhysRevLett.74.4763",
}

@article{shenoy2017quantum,
  title={Quantum cryptography: Key distribution and beyond},
  author={Shenoy-Hejamadi, Akshata and Pathak, Anirban and Radhakrishna, Srikanth},
  journal={Quanta},
  volume={6},
  pages={1--47},
  year={2017},
   doi="https://doi.org/10.12743/quanta.v6i1.57",
}

@article{qureshi2021delayed,
  title={The delayed-choice quantum eraser leaves no choice},
  author={Qureshi, Tabish},
  journal={International Journal of Theoretical Physics},
  volume={60},
  number={8},
  pages={3076--3086},
  year={2021},
  publisher={Springer},
   doi="https://doi.org/10.1007/s10773-021-04906-w",
}

@article{qureshi2020demystifying,
  title={Demystifying the delayed-choice quantum eraser},
  author={Qureshi, Tabish},
  journal={European Journal of Physics},
  volume={41},
  number={5},
  pages={055403},
  year={2020},
  publisher={IOP Publishing},
   doi="https://doi.org/10.1088/1361-6404/ab923e",
}

@misc{qiskit2026,
  author       = {{IBM Quantum}},
  title        = {Qiskit},
  year         = {2026},
  howpublished = {\url{https://www.ibm.com/quantum/qiskit}},
  note         = {Accessed: Aug. 11, 2026}
}

@misc{IBMQuantum2026,
  author       = {{IBM Quantum}},
  title        = {IBM Quantum},
  year         = {2026},
  howpublished = {\url{https://quantum.cloud.ibm.com/}},
  note         = {Accessed: Aug. 11, 2026}
}

@misc{Qniverse2026,
  author       = {{Quantum Technology Group, C-DAC Bengaluru}},
  title        = {Qniverse},
  year         = {2026},
  howpublished = {\url{https://qniverse.in/}},
  note         = {Accessed: Aug. 11, 2026}
}


\end{document}